# Determination of band offsets of Ga$_2$O$_3$/FTO heterojunction for current spreading for high temperature and UV applications

*Carlos G. Torres-Castanedo[1], Kuang-Hui Li[1], Xiaohang Li[1*]*

[1] King Abdullah University of Science and Technology (KAUST), Advanced Semiconductor Laboratory, Thuwal 23955-6900, Saudi Arabia

[2] King Abdullah University of Science and Technology (KAUST), Core Labs, Thuwal 23955-6900, Saudi Arabia

*Corresponding author

**ABSTRACT**

Because of relatively low electron mobility of Ga$_2$O$_3$, it is important to identify proper current spreading materials. Fluorine-doped SnO$_2$ (FTO) offers superior properties to those of indium tin oxide (ITO) including higher thermal stability, larger bandgap, and lower cost. However, the Ga$_2$O$_3$/FTO heterojunction including the important band offset and the I-V characteristics have not been reported. In this work, we have grown the Ga$_2$O$_3$/FTO heterojunction and performed X-ray photoelectron spectroscopy (XPS) measurement. The conduction and valence band offsets were determined to be 0.11 and 0.42 eV, indicating a minor barrier for electron transport and type-I characteristics. The subsequent I-V measurement of the Ga$_2$O$_3$/FTO heterojunction exhibited ohmic behavior. The results of this work manifests excellent candidacy of FTO for current spreading layers of Ga$_2$O$_3$ devices for high temperature and UV applications.

Ultrawide-bandgap semiconductor gallium oxide ($Ga_2O_3$) has the potentials for superior power and optical devices. Different devices using $Ga_2O_3$ have been demonstrated such as MOSFETs,[1] MESFETs,[2] FinFETs,[3] and SBDs.[4] Recently, Green et al.[5] reported a $Ga_2O_3$-based MOSFET with a critical electric field of 3.8 MV/cm, the highest value reported for any transistor. This value is close to half of the theoretical value for $Ga_2O_3$ (8 MV/cm) but already higher than the theoretical limits for GaN (3 MV/cm) and SiC (3.2 MV/cm).[6] $Ga_2O_3$ is also suitable for the solar-blind UV photodetector (SBD) due to its large bandgap (4.7-4.9 eV)[7,8,9] and for gas sensors due to its thermal and chemical stability.[10] For instance, $Ga_2O_3$ thin films have been employed as $O_2$ sensor at high operating temperatures up to 1000 ºC.[11] Moreover, the availability of conductive $Ga_2O_3$ substrates makes this material applicable for vertical injection in visible and UV III-nitride LED technology.[12,13,14]

Ohmic contacts with low contact resistance are essential to accelerate the development of $Ga_2O_3$-based devices. Good p-type doping has not been realized for $Ga_2O_3$[15,16] and thus the discussion of the ohmic contact and the current spreading layer refers to n-type $Ga_2O_3$ only. Recently, the ohmic behavior of nine different metals on n-type $Ga_2O_3$ has been studied, showing that In/Au and Ti/Au form ohmic contact after annealing at 600 and 400-500 ºC, respectively.[17,18] However, the ohmic contacts are not sufficient for high performance $Ga_2O_3$ devices. Because of relatively low electron mobility of $Ga_2O_3$, it is crucial to develop the current spreading layer to reduce current crowding and contact resistance. Recently, Sn-doped indium oxide (ITO) has been studied as a current spreading layer to improve the ohmic contact between metal and $Ga_2O_3$.[19] The conduction and valence band offsets (CBO and VBO) of the $Ga_2O_3$/ITO were recently determined to be 0.32 and 0.78 eV by Carey IV et al, respectively.[20] However, ITO is not an ideal candidate for high temperature and UV applications. First, ITO is thermally unstable at processing or device operation temperatures over 500 ºC.[21, 22, 23] Second, the bandgap of ITO is around 4 eV that makes it absorptive for optical applications below 350 nm. On the other hand, fluorine-doped $SnO_2$ (FTO) could be a better candidate since it is thermally stable even at temperatures higher than 600 ºC.[24] The excellent thermal stability is in particular important for $Ga_2O_3$–based devices as its thermal conductivity is poor which may cause self-heating effects.[25] Moreover, the bandgap of FTO is moderately larger than that of ITO, as measured in this study and presented below, hereby covering a wider range of spectrum in terms of optical transparency. Furthermore, it is worth mentioning that FTO has lower cost than ITO due to scarcity of indium, which can lower overall device cost.[26]

To explore potentials of FTO as the current spreading layer for n-type $Ga_2O_3$, it is essential to identify the band alignment of the $Ga_2O_3$/FTO heterojunction. Ideally, there is no considerable potential barrier for electron transport at the conduction band edge. Furthermore, a non-rectifying electrical behavior would allow FTO to be employed to complement or replace metal contacts, or serve as the current spreading layer. In this study, we report on the band offset measurement of n-type $Ga_2O_3$ grown on commercial FTO substrates by X-ray photoelectron spectroscopy (XPS). The crystal structure and optical transmission of the films were studied by X-ray diffraction (XRD) and UV-Vis spectroscopy. The binding energies and core levels of Ga $2p_{3/2}$ and Sn $3d_{5/2}$ were investigated. The valence and conduction band offsets (VBO and CBO) are determined, where a type-I junction is found. In the end, the Ti/Au metal pads were deposited and annealed to measure the I-V curve of the $Ga_2O_3$:Si/FTO heterojunction. The study paves the way for the use of FTO as the current spreading layer for high temperature and UV applications based on $Ga_2O_3$.

$Ga_2O_3$ thin films have been grown by different techniques such as metalorganic chemical vapor deposition (MOCVD),[2,27] molecular beam epitaxy (MBE),[28] hydride vapor phase epitaxy (HVPE),[29] and pulsed laser deposition (PLD)[30] on both native and foreign substrates. The PLD technique with relatively low cost and high versatility has been employed extensively in the $Ga_2O_3$ research community.[18,31,32]. In this study, specifically, three samples were prepared (Fig. 1), including a commercial 250 nm thick FTO thin film on glass substrate with a sheet resistance of 6 Ω/square (NANOCS FT15-120-20), and 350 and 3 nm thick $Ga_2O_3$:Si thin films deposited by PLD on two FTO/glass substrates, respectively. Prior to PLD, the FTO/glass substrates were sequentially cleaned ultrasonically in acetone and isopropanol, subsequently rinsed in distilled water. Then the 350 and 3 nm $Ga_2O_3$:Si thin films were deposited under the same condition using a Neocera Pioneer 180 PLD system with a chamber base pressure of less than $1\times10^{-7}$ Torr, equipped with a Coherent 205F laser working at 248 nm. A one-inch $Ga_2O_3$ target (PVD Products) with 1.6 at% Si was ablated at a repetition rate of 5 Hz and a pulse energy density of 2 J/cm$^2$. The distance between the target and the substrate was 10 cm. The $O_2$ pressure was 4.5 mTorr and the substrate temperature was 575 °C which is already higher than the ITO stability temperature.

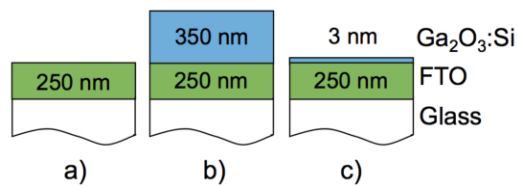

**Figure 1** Schematics of the three investigated samples: (a) the commercial FTO/glass substrate, (b) the thick (350 nm) and (c) the thin (3 nm) $Ga_2O_3$:Si layers deposited on the FTO/glass substrates.

The XPS measurements were carried out immediately after the PLD growth using a Kratos Axis Supra DLD spectrometer with an Al Kα source (λν=1486.6 eV) without any ex-situ cleaning process. The measured binding energies were referenced to the C 1s binding energy of carbon contamination (284.8 eV). The binding energy peaks were fitted by the Voigt curve using a Shirley background subtraction[33)] in the proximity of the peak, while the valence band maximum (VBM) was calculated by extrapolating the leading edge to zero signal. The crystal structure of the thick $Ga_2O_3$:Si and the FTO films was examined by a Bruker D8 Advance X-Ray diffractometer with a Cu Kα source (λ=1.5405Å). The optical transmittance of the films was characterized by a Shimadzu UV-3600 spectrophotometer. Ti (20 nm)/Au (80 nm) contacts were deposited by DC sputtering at 440 W, followed by a rapid thermal annealing (RTA) treatment at 470 °C for 60 s in Ar atmosphere performed in a JetFirst 200C system. The I-V curve of the junction was measured by a Keithley 2400 system.

Fig. 2 shows XRD patterns of the 350 nm $Ga_2O_3$:Si film on the FTO/glass substrate and the FTO/glass substrate itself. These patterns are compared with selected powder peaks of the following JCPDS-ICDD cards: 41-1445 for $SnO_2$ tetragonal-rutile and 43-1012 for $Ga_2O_3$ monoclinic. The FTO/glass substrate is polycrystalline with (101), (110), (200), (211), and (220) planes. Since the substrate is polycrystalline, those skilled in the art should expect the grown thin film to be polycrystalline regardless of the growth or epitaxy techniques. The deposited $Ga_2O_3$:Si film presents a monoclinic structure (β) with two main orientations, predominantly (110) with (400) at much lower intensity (Inset of Fig. 2). These growth directions are consistent with the intensity of (110) and (200) peaks of the FTO film, indicating that $Ga_2O_3$ grew following mainly the (110) plane and marginally the (200) plane of FTO. Due to the monoclinic crystal structure of $Ga_2O_3$, however, the film did not grow in the direction of any of the other three planes, i.e. (101), (211), and (220) of FTO.

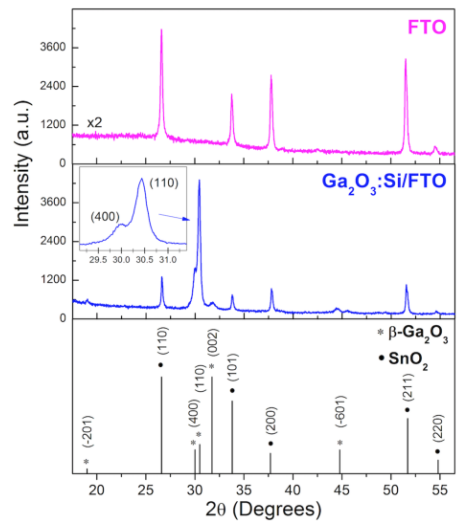

**Figure 2** XRD patterns of the FTO/glass substrate and the 350 nm $Ga_2O_3$:Si film deposited on the FTO/glass substrate. The inset shows the two preferred monoclinic directions for $Ga_2O_3$:Si film in (110) and (400).

To facilitate determination of the band offsets, the bandgap ($E_g$) of $Ga_2O_3$:Si and FTO were firstly deduced from transmission spectra. Since $E_g$ of $Ga_2O_3$ is larger than that of FTO, it is not possible to measure it on the FTO/glass substrate. Thus, a (-201)-oriented 350 nm $Ga_2O_3$:Si layer was grown under the same condition on the optically transparent $c$-sapphire substrate. Fig. 3 shows the transmission measurement for both films using the air baseline while the inset displays the Tauc plot[34] ($h\nu$ vs. $(\alpha h\nu)^n$) in which n=2 was used for directly allowed transitions. The calculated $E_g$'s are 4.94±0.01 eV and 4.40±0.02 eV for $Ga_2O_3$:Si and FTO, respectively. In the case of $Ga_2O_3$:Si, an increment of $E_g$ compared to undoped $Ga_2O_3$ ($E_g$=4.7-4.9eV)[9,10] has been observed concomitantly with the increase of Si content in the film.[35] The $E_g$ of FTO is higher than some reported values (~4.10 eV)[36,37] but it agrees well with the transmission spectra from NANOCS which supplied the FTO substrates for this study.[38]

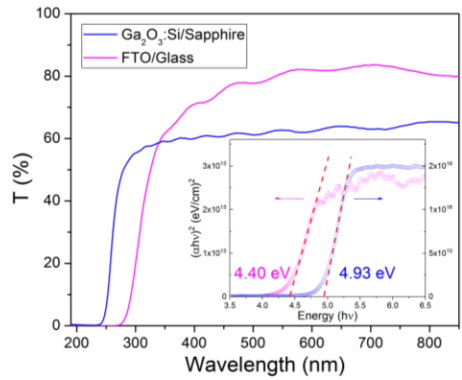

**Figure 3** Transmission spectra of the 350-nm thick $Ga_2O_3$:Si thin films deposited on sapphire and FTO/glass substrates. Inset shows the Tauc plot $h\nu$ vs. $(\alpha h\nu)^2$ with the $E_g$ values for each film.

To determine the band offsets at the heterojunction interface, the Kraut's method[39] was utilized to analyze the XPS spectra of the three samples shown in Fig. 1. First, the core level binding energies and the VBM of the FTO and the 350 nm $Ga_2O_3$:Si layer were determined. The 3 nm $Ga_2O_3$:Si on FTO was measured for the binding energy difference between the two reference core levels at the interface. In all the peak fittings, the Shirley background and Voigt curves were employed. Fig. 4 shows the XPS results. The selected core levels are Ga $2p_{3/2}$ and Sn $3d_{5/2}$ since these are the most intense peaks observed in the XPS survey spectra. The calculation of the VBM for both the $Ga_2O_3$:Si/FTO and FTO is shown in the insets of Fig. 4 a and b, respectively. One Voigt curve allowed the proper fitting of the Ga $2p_{3/2}$ binding energy of both the thick as well as the thin $Ga_2O_3$:Si film on FTO (Fig. 4 a and c). On the other hand, the Sn $3d_{5/2}$ peak is not symmetric due to the contribution of $Sn^{4+}$ and $Sn^{2+}$ (Fig. 4 b and c).[40,41] Table 1 summarizes the core levels and VBM values of the samples.

The VBO and CBO are calculated as follow:

$$\Delta E_V = \left(E_{Ga\,2p}^{Ga_2O_3} - E_{VBM}^{Ga_2O_3}\right) - \left(E_{Sn\,3d}^{FTO} - E_{VBM}^{FTO}\right) + \left(E_{Ga\,2p}^{Ga_2O_3} - E_{Sn\,3d}^{FTO}\right)$$

$$\Delta E_C = \left(E_g^{Ga_2O_3} - E_g^{FTO}\right) - \Delta E_V$$

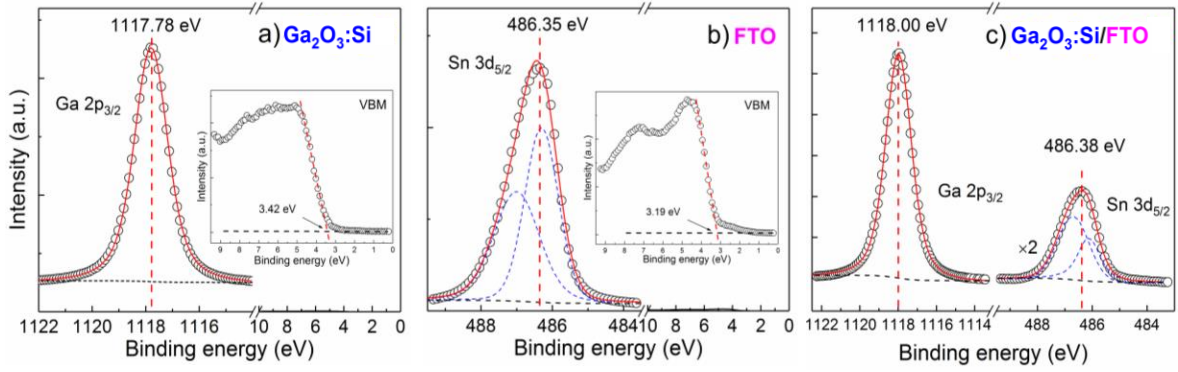

**Figure 4** Core level Ga $2p_{3/2}$ and VBM spectra of the ~350 nm $Ga_2O_3$:Si on FTO (a). Core level Sn $3d_{5/2}$ and VBM spectra of the 250 nm FTO (b). Core level Ga $2p_{3/2}$ and Sn$3d_{5/2}$ of the 3 nm $Ga_2O_3$: Si on FTO (c). The core levels were fitted by Voigt curves, and using the Shirley background.

**Table 1** Peak positions of core levels and VBM used to calculate the band offset in the $Ga_2O_3$:Si/FTO junction.

| Sample | Region | Binding energy (eV±0.10 eV) |
|---|---|---|
| $Ga_2O_3$:Si | Ga 2p 3/2 | 1117.78 |
| | VBM | 3.42 |
| | O 1s | 530.92 |
| FTO | Sn 3d 5/2 | 486.35 |
| | VBM | 3.19 |
| | O 1s | 530.34 |
| $Ga_2O_3$:Si/FTO | Ga 2p 3/2 | 1118.00 |
| | Sn 3d 5/2 | 486.38 |

The diagram in Fig. 5a shows the band alignment diagram of the heterojunction. It shows that the $Ga_2O_3$:Si/FTO junction has a straddling-gap (type-I) alignment, with VBO $\Delta E_V$ of 0.42 eV and CBO $\Delta E_C$ of 0.11 eV. Since the $\Delta E_C$ is small, this alignment is desirable for electron transport across the heterointerface. Previously, a type-I junction was reported for polycrystalline [(-401), (-601)] $Ga_2O_3$ deposited by PLD on a (111) Si substrate with $\Delta E_C$ as low as 0.2 eV.[42] In another study, the band offset of ITO deposited by sputtering (ITO's crystal structure was not reported) on (-201) $Ga_2O_3$ led to $\Delta E_C$=0.32 eV and a type-I junction.[22] Our results show twice and three times lower $\Delta E_C$ than these studies which favor the electron transport. Based on our exhaustive literature survey, there has not been any report or evidence showing that the band alignment can be altered by changing the doping level or polycrystalline grain boundary. Thus we do not draw any hypothesis and conclusion here. In addition, there may be strain in the $Ga_2O_3$ layer due to lattice mismatch. However, previous works have shown that the impact of strain on CBO and VBO is nearly negligible by comparing the unstrained and strained heterojunctions.[43,44]

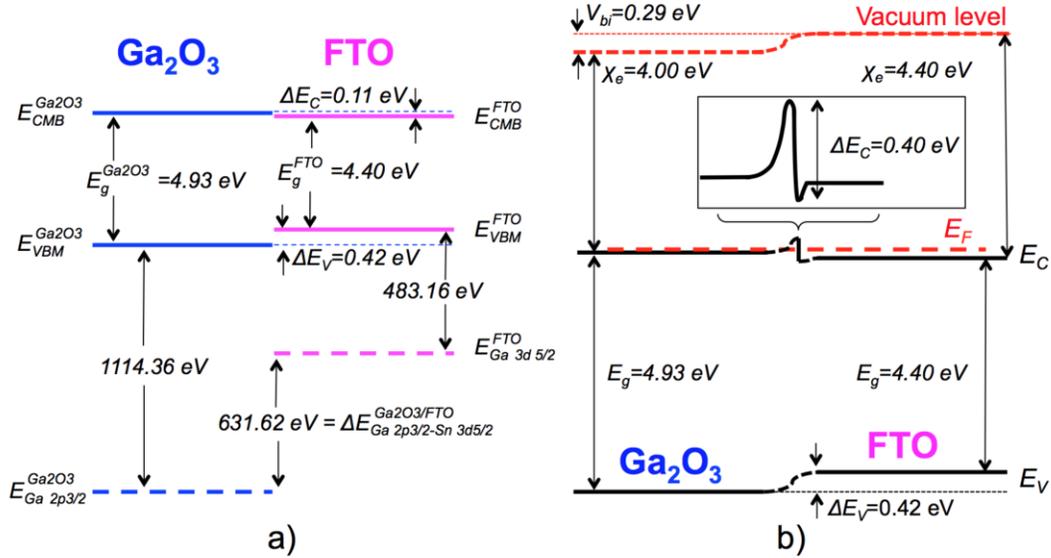

**Figure 5** (a) Band alignment diagram for the $Ga_2O_3$/FTO heterojunction obtained by XPS and (b) the band diagram schematic with the band bending in the junction at RT.

Furthermore, the band bending for the $Ga_2O_3$:Si/FTO heterojunction is shown in Figure 5b, aligning the Fermi level of both materials. The effective density of states function in the conduction band for both materials was calculated considering an effective electron mass ($m_n^*$) of 0.28 $m_o$ corresponding to $SnO_2$ and $Ga_2O_3$.[45,46] According to our Hall effect measurements at RT, the concentrations of electrons were $9.1 \times 10^{20}$ and $1.0 \times 10^{19}$ cm$^{-3}$ for FTO and $Ga_2O_3$:Si films, respectively. Hence, the Fermi levels for the materials located above the conduction band (0.14 eV for FTO and 0.03 eV for $Ga_2O_3$). Considering the calculated Fermi levels and the reported electron affinities ($\chi_e$) for $Ga_2O_3$ (4.0 eV) and FTO (4.4 eV)[47,48], a built-in potential ($V_{bi}$) of 0.29 eV was obtained. A type-I junction is still observed with the same $\Delta E_V$ but a higher $\Delta E_C$ compared to the band alignment study occurs due to band bending. To investigate the electrical properties of the $Ga_2O_3$/FTO and FTO/metal heterojunctions, we performed I-V measurement at RT. The I-V curve and the schematic are presented in Fig. 6. The I-V curve was measured before and after rapid thermal annealing (RTA) at 470 ºC in Ar atmosphere for the Ti/Au contacts. After annealing, the resistance decreased significantly from approximately 27 to 12 Ω while the ohmic behavior was observed, indicating that FTO can be an excellent current spreading layer for $Ga_2O_3$.

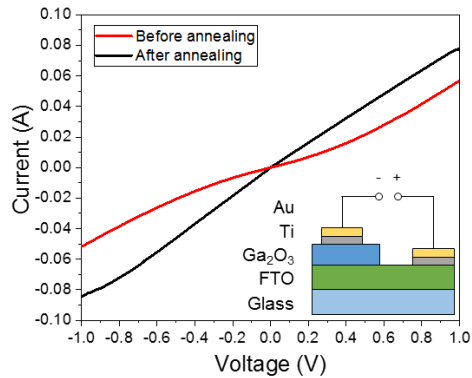

**Figure 6** I-V curves of the $Ga_2O_3$:Si/FTO junction before and after the annealing process for the Ti/Au contacts. Inset shows the cross-sectional schematics.

In summary, we reported on formation and characterization of the $Ga_2O_3$/FTO heterojunction. In particular, we have performed high-resolution XPS measurements to determine that $Ga_2O_3$:Si/FTO heterojunction has a straddling-gap (type-I) alignment with $ΔE_V$ of 0.42±0.10 eV and $ΔE_C$ of 0.11±0.10 eV. The junction exhibits an ohmic behavior with Ti/Au contacts after annealing. The small $ΔE_C$ and the non-rectifying behavior of the junction, as well as the large $E_g$ of 4.40 eV and thermal stability at high temperature, make FTO a promising candidate for use as a current spreading layer in $Ga_2O_3$-based high temperature and short wavelength devices.

The authors would like to acknowledge the support of KAUST Baseline BAS/1/1664-01-01, KAUST Competitive Research Grant URF/1/3437-01-01 and URF/1/3771-01-01, and GCC Research Council REP/1/3189-01-01. Also, we thank Prof. Rebeca Castanedo Perez from the Materials Science Department of the Center for Research and Advanced Studies of the National Polytechnic Institute for providing the FTO/glass substrates.